\documentclass[useAMS,usenatbib,referee]{biom}

\usepackage{amsmath}
\usepackage{amssymb}
\usepackage{latexsym}
\usepackage{graphics, graphicx}
\usepackage{epsfig,epsf,rotating}
\usepackage{epsf}
\usepackage{theorem}
\usepackage{bm}
\usepackage{booktabs}

\usepackage{longtable}
\usepackage[caption=false]{subfig}
\usepackage{cleveref, autonum}
\usepackage[ruled]{algorithm2e} 

\newcommand{\I}{\text{I}}

\usepackage[usenames,dvipsnames]{color}
\usepackage{color,soul}
\usepackage{xcolor,framed}
\definecolor{shadecolor}{HTML}{FFF200}
\usepackage{ifthen}
\newboolean{isclean}
\setboolean{isclean}{false}
\ifthenelse{\boolean{isclean}}{

}{}

\usepackage{mydefs_new}

\newcommand{\bstheta}{\boldsymbol{\theta}}

\newcommand{\mc}[1]{\mathcal{#1}}

\newcommand{\iid}{\stackrel{\rm iid}{\sim}}

\renewcommand{\Cov}{\text{cov}}
\renewcommand{\cov}{\Cov}

\RequirePackage[OT1]{fontenc}
\RequirePackage{etex}
\usepackage{mathrsfs}

\title{Bayesian Inference for Stationary Points in Gaussian Process Regression Models
for Event-Related Potentials Analysis}

\author{Cheng-Han Yu$^1$, Meng Li$^2$, Colin Noe$^3$, Simon Fischer-Baum$^3$ and Marina Vannucci$^2$
\email{marina@rice.edu}\\
$^1$Department of Mathematical and Statistical Sciences, Marquette University, Milwaukee, WI, USA\\
$^2$Department of Statistics, Rice University, Houston, TX, USA\\
$^3$Department of Psychological Science, Rice University, Houston, TX 77005.
}

\begin{document}

 \date{{\it Received ...} .... {\it Revised ...}.\newline 
{\it Accepted ...} }


\volume{...}
\artmonth{...}
\doi{10.1111/j.1541-0420.2005.00454.x}


\label{firstpage}

\begin{abstract}
Stationary points embedded in the derivatives are often critical for a model to be interpretable and may be considered as key features of interest in many applications. We propose a semiparametric Bayesian model to efficiently infer the locations of stationary points of a nonparametric function, which also produces an estimate of the function. We use Gaussian processes as a flexible prior for the underlying function and impose derivative constraints to control the function's shape via conditioning. We develop an inferential strategy that intentionally restricts estimation to the case of at least one stationary point, bypassing possible mis-specifications in the number of stationary points and avoiding the varying dimension problem that often brings in computational complexity.  We illustrate the proposed methods using simulations and then apply the method to the estimation of event-related potentials (ERP) derived from electroencephalography (EEG) signals. We show how the proposed method automatically identifies characteristic components and their latencies at the individual level, which avoids the excessive averaging across subjects which is routinely done in the field to obtain smooth curves. By applying this approach to EEG data collected from younger and older adults during a speech perception task, we are able to demonstrate how the time course of speech perception processes changes with age.\\
\end{abstract}

\begin{keywords}
Bayesian inference; Semiparametric; Gaussian process; Derivatives; ERP data.
\end{keywords}

\maketitle

\section{Introduction}
Shape constraints in the form of  stationary points embedded in the derivatives arise naturally in a wide range of modern applications. Estimation of the stationary points of functional curves, in addition to the fitted curves, may be considered a key feature of interest in many applications. In our motivating example, event-related potentials (ERPs) are derived by averaging electroencephalography (EEG) signals, collected in response to specific stimuli or events, and used to infer brain electrical potentials \citep{Gasser1996}.  ERP waveforms consist of characteristic components that span across time, and that relate to specific mental or cognitive processes, such as signal matching, decision making, language processing and memory updating \citep{Luck2012}. Analyses of ERP data often focus on identifying specifically meaningful peaks, or characteristic components, and computing 
their latency (time when the peak or dip occurs) and/or amplitude (magnitude of the peak) \citep{Luck2005}. The two most common methods are hand selecting peaks or integrating the area under the curve over a pre-set time-window, and these methods have remained largely unchanged over the past fifty years  \citep{Luck2012}.  In recent years, some alternative approaches have been proposed for detecting ERPs in single subjects, such as those based on the studentized continuous wavelet transformation technique, see for example \cite{Kalli} for discussion and a comparison of these techniques. However, these methods are typically designed to detect whether or not an ERP component is present, either for the development of brain-computer interface devices or for assessment of clinical characteristics, such as the person’s state of consciousness. This research therefore differs from the goal of our work, which is the estimation of stationary points of ERP curves at the subject level.
In statistics, there is limited literature on ERP data; a notable exception is the moving average-based meta-preprocessing procedure of \cite{hasenstab2015identifying,hasenstab2017multi}, to increase signal-to-noise ratio. Our focus is instead on detecting stationary points under noisy ERP data.

We propose a semiparametric Bayesian model to efficiently infer the locations of stationary points of a nonparametric function, which also produces an estimate of the function. Specifically, we use Gaussian processes (GPs) as a flexible prior for the underlying function and impose derivative constraints to control the function's shape. This leads to a new process bearing a particular form of GPs, which is referred to as \textit{Derivative-constrained GP}, as it incorporates derivative constraints via conditioning.  GPs have been arguably one of the most widely used nonparametric processes for continuous curves in Bayesian statistics and machine learning, partly due to their flexibility and tractability, see \cite{rasmussen2006} and \cite{Ghosal+van:17} for comprehensive treatments. \cite{Holsclaw2013} estimate the derivative of a curve using a GP-based inverse method.
\cite{Zhou2019} use constrained Gaussian processes to enforce constraints on the parameter space such as convexity and inequality of linear combinations through indicator functions. \cite{Wang2016} study various strategies including constrained Gaussian processes and conditional Gaussian processes by conditioning an unrestricted Gaussian process on a positive probability event. While the emphasis in these approaches is to investigate the benefit of adding shape constraints to the fitting of the original function, in this paper we are primarily interested in the estimation of the stationary points of the underlying function. 

In our application to EEG data, the derivative constraints are only partially specified, that is, the prior knowledge is only concerned with the existence of finitely many stationary points, while their number and locations are left unknown.  We adopt the conditioning approach by unifying the curve fitting and stationary points detection, which is presumably more suited for extremely low signal-to-noise ratio data such as EEG.  This allows us to intentionally restrict estimation to the case of \textit{at least one stationary point}, bypassing possible mis-specifications in the number of stationary points and avoiding the varying dimension problem that often brings in computational complexity. For posterior inference, we propose a Monte Carlo expectation-maximization (MCEM) algorithm that includes hyperparameters selection.
We illustrate the proposed method using simulations and then show an application to ERP data derived from EEG signals, where we estimate smooth curves and automatically identify characteristic components and their latencies through the detection of the stationary points of the curves.  By applying our approach to data collected from younger and older adults during a speech perception task, we are able to demonstrate how the time course of speech perception processes changes with age. We explicitly account for error in the data and produce estimate at the individual level, avoiding the excessive averaging across subjects routinely done to obtain smooth curves.  
This provides a major advance over standard approaches in the clinical and cognitive literature on ERP data, where there is increasing interest in relating individual differences in ERP components to behavior  \citep{Kim, Tanner, Hajcak}. Furthermore, unlike exploratory methods that rely on visual inspection of the curves, our method provides uncertainty quantification on the estimated quantities via high posterior density regions. 

Section \ref{sec:methods} presents model, priors and posterior inference, Section \ref{sec:simul} simulations, Section \ref{sec:ERP} the case study and Section \ref{sec:disc} concluding remarks. 

\section{Methods}
\label{sec:methods}

\subsection{Shape-constrained regression} 
Let $\bm{y} = \{y_1, \ldots, y_n\}$ be noisy observations from an unknown function $f: \mathcal{X} \rightarrow \mathbb{R}$ observed at locations $\bm{x} = \{x_1, \ldots, x_n\}$. We assume a nonparametric regression model of the type
\begin{eqnarray}
y_i = f(x_i) + \epsilon _i, ~~ i = 1, \dots, n, 
\label{eq:data_generating}
\end{eqnarray}
where $\epsilon_i$ is Gaussian random noise, i.e., $\epsilon_i \iid N(0, \sigma ^ 2)$. We are interested in situations where $f(\cdot)$ has $M$ stationary points, 
\begin{equation}
\label{eq:1st_der}
f'(t_m) = 0, ~~m = 1, \dots, M,
\end{equation}
with $f'(\cdot)$ indicating the first derivative of $f(\cdot)$.  Model construction \eqref{eq:data_generating}--\eqref{eq:1st_der} extends the classic nonparametric regression model to incorporate derivative constraints on the unknown regression function.  The stationary points $\bm{t} = (t_1, \ldots, t_M)$ in \eqref{eq:1st_der}, which may or may not coincide with the input values in \eqref{eq:data_generating}, are our parameters of interest, while the underlying function $f(\cdot)$ is a nuisance parameter. One primary objective is to estimate the number of stationary points $M$ and their locations $\bm{t}$. Below we propose a novel Bayesian method to fit the model and carry out statistical inference. In particular, we use derivative-constrained Gaussian processes (DGP) to incorporate the shape constraints on $f(\cdot)$ and intentionally restrict to the case of \textit{at least one stationary point}, to bypass the specification of a prior on $M$ and avoid the varying dimension of $\bm{t}$ that often brings in computational complexity. We also derive inferential tools to quantify the uncertainties in the estimation of $\bm{t}$. 

\subsection{Derivative-constrained Gaussian processes (DGP)} \label{section:DGP} 
We model the unknown function $f(\cdot)$ via a Gaussian process prior 
\begin{eqnarray}
f(\cdot) \sim GP(\mu(\cdot), k(\cdot, \cdot)), 
\label{eq:GP}
\end{eqnarray}
with mean function $\mu(x_i) = \E(f(x_i))$ and covariance function 
$k(x_i, x_j) =\Cov(f(x_i), f(x_j))$. 
For concreteness, we focus here on $\mc{X} = [a, b]$ and notice that the proposed methods are readily applicable for general domains such as $\mc{X} = [a, b]^d$ for $d \geq 1$. 
It is well known that a GP possesses a version whose sample paths are differentiable if its covariance kernel is continuously differentiable~\citep[page 574]{Ghosal+van:17}; throughout this paper, we always refer to such differentiable sample paths for a GP. 
Our approach utilizes the desirable property that the derivative of a GP, along with the original sample path, is also jointly a GP, provided that the GP is mean-square differentiable; see for example Ch. 2 of \cite{Adler1981}. This property makes Gaussian processes well suited for our purpose to study the location of the stationary points when the true function is unknown.

\begin{figure}
	\centering
	\includegraphics[width=\textwidth]{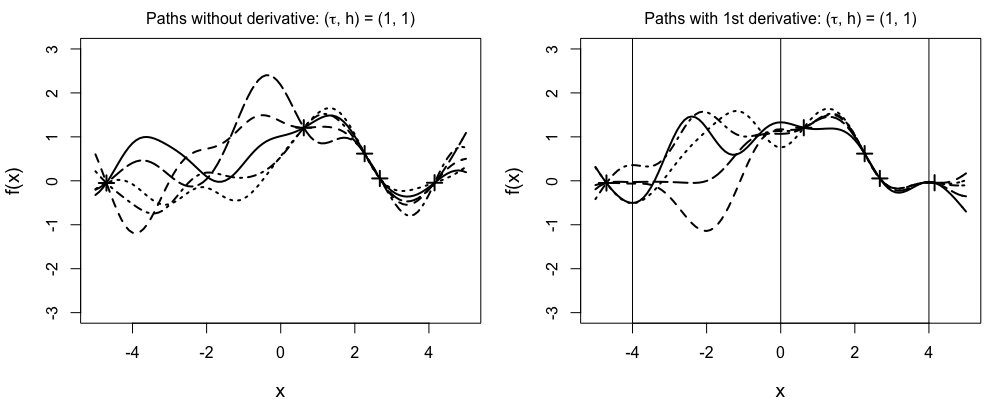}
	\caption{Sample paths from a Gaussian process with (right) and without (left) first order derivative information. A squared exponential kernel \eqref{eq:SEcov} with parameters $\tau=1$ and $h=1$ was used.  The left-hand panel shows sample paths from a standard GP with input values randomly chosen from $[-5, 5]$, indicated with plus signs in the plot, while the right-hand panel shows sample paths from a DGP with stationary points $t_1 = -4, t_2 = 0,$ and $t_3 = 4$, indicated with vertical lines in the plot. As expected, when the derivative constraints are added to the process, all curves have derivatives equal to zero at $t_m, m = 1, 2, 3$, but each curve generally takes its own values at $t_m, m= 1, 2, 3$. }
	\label{fig:paths}
\end{figure}

We indicate with $W = \{W_x, x \in \mc{X}\}$ a differentiable sample path of $GP(\mu(\cdot), k(\cdot, \cdot))$.
Given two arbitrary points $(x, \tilde{x})$ from $\mc{X}$, we have 
\begin{align} \label{eq:cov_f_fd}
\cov\left(W'_{\tilde{x}},W_x\right) = \frac{\partial}{\partial \tilde{x}}\cov \left(W_{\tilde{x}},W_x\right) & = \frac{\partial}{\partial \tilde{x}}k\left(\tilde{x}, x\right) := k_{10}(\tilde{x}, x)  \\
\cov\left(W_x, W'_{\tilde{x}}\right) = \frac{\partial}{\partial \tilde{x}}\cov \left(W_x, W_{\tilde{x}}\right) & = \frac{\partial}{\partial \tilde{x}}k\left(x, \tilde{x}\right) := k_{01}(x, \tilde{x}) = k_{10}(\tilde{x}, x)\\
\cov\left(W'_{x}, W'_{\tilde{x}}\right) = \frac{\partial ^ 2 }{\partial x \partial \tilde{x}} \cov \left(W_{x}, W_{\tilde{x}}\right) & = \frac{\partial ^ 2 }{\partial x \partial \tilde{x}} k\left(x, \tilde{x}\right) := k_{11}(x, \tilde{x}). 
\end{align} 
Extending the derivative operation to two arbitrary vectors $\bm{x} = (x_1, \ldots, x_J)$ and $\tilde{\bm{x}} = (\tilde{x}_1, \ldots, \tilde{x}_{\tilde{J}})$ yields a random vector $(W_{x_1}, \ldots, W_{x_J}, W'_{\tilde{x}_1}, \ldots, W'_{\tilde{x}_{\tilde{J}}})$ which is distributed according to a multivariate normal distribution with mean $(\mu(\bm{x})^T, \mu'(\tilde{\bm{x}})^T)^T$ and covariance  
\begin{equation}
\begin{bmatrix}
k(\bm{x}, \bm{x}) & k_{01}(\tilde{\bm{x}}, \bm{x}) \\
k_{01}(\bm{x}, \tilde{\bm{x}})  &  k_{11}(\tilde{\bm{x}}, \tilde{\bm{x}}) 
\end{bmatrix}, 
\end{equation}
where we adopt the shorthand $\mu(\bm{x})$ to denote the vector $(\mu(x_1), \ldots, \mu(x_J))$, $k_{10}(\tilde{\bm{x}}, \bm{x})$ to denote the matrix whose $ij$th entry is $k_{10}(\tilde{x}_i, x_j)$, with the same convention for $\mu'(\tilde{\bm{x}})$, $k(\bm{x}, \bm{x})$, $k_{01}(\bm{x}, \tilde{\bm{x}})$, and $k_{11}(\tilde{\bm{x}}, \tilde{\bm{x}})$. Consequently, the conditional distribution of $(W_{x_1}, \ldots, W_{x_J})$ given $(W'_{\tilde{x}_1}, \ldots, W'_{\tilde{x}_{\tilde{J}}}) = \bm{0}$ is a multivariate normal with mean
$[\mu(\bm{x}) - k_{01}(\bm{x}, \bm{\tilde{x}}) k_{11}^{-1}(\tilde{\bm{x}}, \tilde{\bm{x}}) \mu'(\tilde{\bm{x}})]$ and covariance 
$[k(\bm{x}, \bm{x}) - k_{01}(\bm{x}, \tilde{\bm{x}}) k_{11}^{-1}(\tilde{\bm{x}}, \tilde{\bm{x}}) k_{10}(\tilde{\bm{x}}, \bm{x})]$.
This leads to the following definition of a Gaussian process with derivative constraints at stationary points $\bm{t} = (t_1, \ldots, t_M)$.
\begin{definition}\label{def1}
A random process $W$ is said to be a derivative-constrained Gaussian process at points $\bm{t}$, denoted by DGP$(\mu, k, \bm{t})$,  if it is a Gaussian process with mean $x \mapsto \mu(x) - k_{01}(x, \bm{t}) k_{11}^{-1}(\bm{t}, \bm{t}) \mu'(\bm{t})$
	and covariance kernel 
$(x, \tilde{x}) \mapsto [k(x, \tilde{x}) - k_{01}(x, \bm{t}) k_{11}^{-1}(\bm{t}, \bm{t}) k_{10}(\bm{t}, \tilde{x})]$.
\end{definition} 
Sample paths $W$ of a DGP$(\mu, k, \bm{t})$ have zero derivatives at $\bm{t}$ almost surely; this is because $\E(W'(\bm{t})) = \mu'(\bm{t}) - k_{11}(\bm{t}, \bm{t}) k_{11}^{-1}(\bm{t}, \bm{t}) \mu'(\bm{t}) = \bm{0}$, and covariance $\cov(W'(\bm{t}), W'(\bm{t})) = k_{11}(\bm{t}, \bm{t}) - k_{11}(\bm{t}, \bm{t}) k_{11}^{-1}(\bm{t}, \bm{t}) k_{11}(\bm{t}, \bm{t}) = \bm{0}$.
We note that, unlike most conditional GPs used in the literature, the conditioning event in our DGP has zero probability and involves unknown parameters.
In the sequel, we use zero mean $\mu(\cdot) = 0$, and a \textit{squared exponential} (SE) kernel
\begin{equation}
k(x_i, x_j) = \tau^2 \exp\left(-\frac{1}{2h^2}\|x_i - x_j\|^2\right), 
\label{eq:SEcov}
\end{equation}
which is infinitely differentiable. The parameter $\tau^2$ controls the vertical variation of the process, and $h$ is the so called length scale parameter that controls the correlation range of the process. A larger $h$ results in higher correlation between inputs and ends up with a smoother curve. The induced DGP is defined based on the partial derivatives of $k(\cdot, \cdot)$: 
\begin{align}
k_{01}\left(x_i, t_j\right) & = k_{10}(t_j, x_i) = \tau^2 \exp\left(-\frac{1}{2h^2}|x_i - t_j|^{2}\right)\frac{\left(x_i - t_j\right)}{h^2},\\
k_{11}\left(t_i, t_j\right) & = \tau^2 \exp\left(-\frac{1}{2h^2}|t_i - t_j|^{2}\right) \left(\frac{1}{h^2}\left(1 - \frac{\left(t_i - t_j\right)^2}{h^2} \right)\right). 
\end{align}
Other kernels with required differentiability can also be used; see \cite[Ch 4]{rasmussen2006} for a detailed discussion of covariance functions. Figure \ref{fig:paths} shows some sample paths from a Gaussian process obtained with the SE kernel \eqref{eq:SEcov} and parameters $\tau=1$ and $h=1$, with and without derivatives constraints.  It is important to notice that imposing derivative constraints on $M$ points implies that every GP path will have {\it at least} $M$ stationary points. Below we exploit this simple fact when imposing priors on the stationary points.

\subsection{Priors on stationary points}
Let's consider model \eqref{eq:data_generating}--\eqref{eq:1st_der}, with the GP prior \eqref{eq:GP} on $f$. 
A standard strategy to specify the prior $\pi(\bm{t})$ on the stationary points is to specify a hierarchical distribution through an $M$-dimensional prior for $\bm{t}$ supported on $\mc{X}^M$, with a hyperprior on $M$. This strategy inevitably involves a varying dimension of $\bm{t}$ in the posterior sampling and model selection for various values of $M$, and a reversible jump Markov chain Monte Carlo strategy for sampling purposes.
Alternatively, we propose to use a univariate stationary point $t$ with a prior $\pi(t)$ supported on $\mc{X}$, which corresponds to assuming that the regression function has {\it at least one} stationary point, and then utilize the posterior of $t$ to infer all stationary points. This strategy, which we refer to as \textit{single-DGP}, bypasses reversible jump, essentially relying on a multi-modal posterior of $t$. Later we show results on simulated and real data that support this claim.
As for the prior on the univariate $t$, we use the Beta distribution. We complete the prior specification by imposing a standard inverse gamma prior $IG(a_{\sigma},  b_{\sigma})$ on $\sigma^2$.

\subsection{Posterior inference via Monte Carlo EM}
\label{sec:MCEM}
For posterior inference, we develop a Monte Carlo expectation-maximization (MCEM) approach \citep{wei1990}. In particular, we adopt the data-adaptive algorithm of \cite{Bai2019} that embeds the EM algorithm for estimating the hyperparameters based on marginal maximum likelihood (MML) into the posterior simulation updates, therefore avoiding separate tuning of the hyperparameters.

We reparameterize the hyperparameters in the SE kernel $(\tau, h)$ as $\bstheta = (\tau_0, h)$ with $\tau = \tau_0 \sigma$. At a generic iteration, in the Monte Carlo E-step we fix $\bstheta$ at its current value $\hat{\bstheta}^{(i)}$ and sample $t$ via standard  Metropolis-Hastings \citep{MH70} with a uniform proposal and $\sigma^2$ from its inverse-gamma full conditional distribution. In the M-step, we estimate the hyperparameters $\bstheta$ based on the MML \citep{Bai2019}. With $f$ being marginalized out, such $\bstheta$ gives the maximal model evidence and most likely kernel structure. Suppose we obtain a set of $D$ samples from the posterior of $t$ and $\sigma^2$, i.e.  $\{t_1, \dots, t_D\}$ and $\{\sigma_1^2, \dots, \sigma_D^2\}$, from the E-step. In the M-step, following \cite{Bai2019}, we sample $\{t_j\}_{j = 1}^J$ and $\{\sigma_j^2\}_{j = 1}^J$ with $J < D$ to be used to optimize $\bstheta$. The conditional log-likelihood on $\bstheta$ and $\bm{y}$ in the E-step is given by $$Q(\bstheta \mid \hat{\bstheta}^{(i)}) = \E_{t, \sigma^2 \mid \hat{\bstheta}^{(i)}, \bm{y}}[ \log p\left(\bm{y} \mid t, \sigma^2, \bstheta \right)]$$ and the MML estimate $\hat{\bstheta}^{(i+1)}$ is 
\begin{eqnarray}
\label{mstep_mml}
\hat{\bstheta}^{(i+1)} = \arg \max_{\bstheta} \hat{Q}(\bstheta \mid \hat{\bstheta}^{(i)}) = \frac{1}{J} \sum_{j=1}^J \log  p\left(\bm{y} \mid t_j^{(i)}, (\sigma^2_j)^{(i)}, \hat{\bstheta}^{(i)} \right),
\end{eqnarray}
which can be obtained numerically. The proposed MCEM algorithm is detailed in Algorithm \ref{algo:StoEM}. Convergence of the algorithm is reached when $\| \hat{\bstheta}^{(i+1)} - \hat{\bstheta}^{(i)} \| < \epsilon$ at the $(i+1)$-th iteration. \cite{Bai2019} recommend setting the threshold $\epsilon$ to a small value and then use  $J=200$ and $D=20000$, which they chose essentially by trial and error, to achieve a balance between performance and computation time. In this paper, we used $J=500$ and $D=5000$ with $\epsilon$ set to 0.0001, in all simulations and real data analyses. We also used the Gelman and Rubin’s convergence diagnostic \citep{gelman1992inference} on the posterior samples drawn in the E-step to check for signs of non-convergence of the individual parameters. Those statistics were all below 1.1, clearly indicating that the MCMC chains were run for a sufficient number of iterations.

\begin{algorithm}
	\fontsize{12}{6}\selectfont
	\SetAlgoLined
	\textbf{Initialize:}\\
    Set values of $\epsilon$, $D$, $J$, $a_{\sigma}$, $b_{\sigma}$\\
	Set initial value $\hat{\bstheta}^{(1)}$, $t_0$ and $\sigma_0^2$.\\
	\vskip 2mm
	For iteration $i = 1, 2, \dots$ until convergence criterion $\|\hat{\bstheta}^{(i)} - \hat{\bstheta}^{(i+1)}\| < \epsilon$ is satisfied:\\
	\vskip 2mm
	\textbf{Monte Carlo E-step:}\\ 
	\vskip 1mm
	\eIf{$i = 1$}{
		$\{t_d^{(i)}\}_{d = 1}^D \sim Unif[a, b]$\;
	}{
    \For{$d=1, \dots, D$}{
        \textbf{Metropolis-Hastings for $t$}\\
		$\left(t_d^{(i)}\right)^* \sim Unif[a, b]$, i.e., the proposal is a uniform distribution\;
		$u \sim Unif(0, 1)$\;
		\eIf{$u < \pi\left(\left(t_d^{(i)}\right)^* \mid \bm{y},(\sigma_d^2)^{(i)}, \hat{\bstheta}^{(i)}\right) / \pi\left(t_{d-1}^{(i)} \mid \bm{y},(\sigma_d^2)^{(i)}, \hat{\bstheta}^{(i)}\right)$}{$t_d^{(i)} = \left(t_d^{(i)}\right)^*$}{$t_d^{(i)} = t_{d-1}^{(i)}$}
		\textbf{Gibbs for $\sigma^2$}\\
		$(\sigma_d^2)^{(i)} \sim IG\left(\frac{n}{2} + a_{\sigma}, \frac{1}{2}\bm{y}^TA^{-1}(t_d^{(i)})\bm{y} + b_{\sigma} \right)$\; with $A(t) = \tau_0^2 \left( k(\bm{x}, \bm{x}) - k_{01}(\bm{x}, t) k_{11}^{-1}(t, t) k_{10}(t, \bm{x}) + \I_n\right)$

		$d \leftarrow d + 1$}
	}
	\vskip 2mm
	Draw sample $\{t_j^{(i)}\}_{j = 1}^J$ from $\{t_d^{(i)}\}_{d = 1}^D$.\\
	Draw sample $\{(\sigma_j^2)^{(i)}\}_{j = 1}^J$ from $\{(\sigma_d^2)^{(i)}\}_{d = 1}^D$. 
	\vskip 2mm
	\textbf{M-step:}\\ Given samples of $t$ and $\sigma^2$ at the $i$-th iteration,  $\{t_j^{(i)}\}_{j = 1}^J$ and $\{(\sigma_j^2)^{(i)}\}_{j = 1}^J$, update $\bstheta$ to $\hat{\bstheta}^{(i+1)}$ by optimizing
	$\hat{\bstheta}^{(i+1)}$ according to (\ref{mstep_mml}).
	\caption{Monte Carlo EM Algorithm for single-DGP}
	\vskip 2mm
	\KwResult{MML estimate of $\bstheta$ and posterior samples of $t$ and $\sigma^2$}
	\label{algo:StoEM}
\end{algorithm}

Point estimates of the stationary points can be obtained from the posterior distributions of $t$. Here we use as posterior summary the highest posterior density (HPD) interval, that not only provides a Bayesian credible interval for $t$ but also gives a natural way to estimate the number of stationary points $M$ as the number of discrete segments that compose the HPD interval. Empirical estimates of the HPD interval for $t$ can be calculated as follows: Given the posterior samples of $t$, one can obtain the estimated density of $t$, say $g(t)$, and then the $100(1-\alpha)\%$ HPD interval is the subset $C(g_{\alpha})$ of $\mc{X}$ such that $C(g_{\alpha}) = \left\{t:g(t) \geq g_{\alpha} \right\}$ where $g_{\alpha}$ is the largest constant such that $P\left(t \in C(g_{\alpha})\right) \geq 1 - \alpha$.
Given the HPD interval, an estimate of the $j$-th stationary point is derived as the maximum a posteriori (MAP) estimate calculated as the mode of the $j$-the segment identified by the HPD interval.

Our inferential strategy produces estimates not only of the stationary points, but also of the unknown regression function $f$. Let $\{t_d, \sigma^2_d\}_{d=1}^D$ be final posterior samples of $t$ and $\sigma^2$, and $\tau^*$ and $h^*$ the MML estimates of $\tau$ and $h$. Given any grid inputs $\bm{x}^*$, we sample the $d$-th sample path $f$ at $\bm{x}^*$ from the posterior distribution $\pi\left(f \mid t_d, \sigma^2_d, \tau^*, h^*, \bm{y},\bm{x}^* \right)$ for $d = 1, 2, \dots, D$, which is a multivariate normal $N \left(\bm{0}, \B\right)$, where 
$$
\B = k(\bm{x}^*, \bm{x}^*) -  
\begin{bmatrix} 
k(\bm{x}, \bm{x}^*) \\
k_{10}(\bm{t}, \bm{x}^*)
\end{bmatrix}^T
\begin{bmatrix}
k(\bm{x}, \bm{x}) & k_{01}(\bm{x}, \bm{t}) \\
k_{10}(\bm{t}, \bm{x}) & k_{11}(\bm{t}, \bm{t})
\end{bmatrix}^{-1}\begin{bmatrix} 
k(\bm{x}, \bm{x}^*) \\
k_{10}(\bm{t}, \bm{x}^*)
\end{bmatrix}.
$$

\section{Simulation Study} \label{sec:simul} 
In this section, we examine the performance of our proposed method via a simulation study.  We consider a basic univariate response, with a one dimensional covariate $x$, and a nonlinear regression function
$f(x) = (0.3 + 0.4x + 0.5\sin(3.2x) + 1.1/(1+x^2))$,
with $x$ in $\mc{X} = [0, 2]$ and two stationary points at $t_1 = 0.436$ and $t_2 = 1.459$. We set $n = 50$ and generate 100 independent data sets according to model \eqref{eq:data_generating}--\eqref{eq:1st_der} with $\sigma = 0.25$ and inputs $\{x_i\}_{i = 1}^n$ generated from a $Unif(0, 2)$ distribution. 
Results we report were obtained by fitting DGP models with zero mean and SE kernel \eqref{eq:SEcov}. When fitting our {\it single-DGP} method, we imposed a $Unif(0, 2)$ prior on $t$ and a vague $IG(1/2, 1/2)$ prior on $\sigma^2$. 
Although not of primary interest in our motivating case study, the estimation of the regression function is also considered in the simulations, with an equally-spaced test grid, $\{x^*_j\}_{j = 1}^{100}$ used for visualization. 

Our {\it single-DGP} setting treats the location of the stationary points as random and unknown and, furthermore, assumes that there is one stationary point only, $t$. As previously remarked, imposing one stationary point results in curves with at least one stationary point. This strategy, therefore, avoids the issue of mis-specifying the number of stationary points by letting the model learn the number and their locations. In this simulation, we compare performances of single-DGP with two additional settings. In one setting we assume that one has prior knowledge about the number of stationary points and some knowledge about their locations. In particular, here we assume two stationary points, $t_1$ in the interval $(0, 1)$ and $t_2$ in $(1, 2)$. We then assign prior distributions $t_1\sim Unif(0, 1)$ and $t_2\sim Unif(1, 2)$. We call this setting {\it multiple-DGP}. In the Supplementary Material, we provide a description of an adaptation of our MCEM algorithm to this multiple-DGP framework. 
We also consider an {\it Oracle-DGP} setting, which assumes the number of stationary points and their locations known and incorporated into the model via the constraint $f'(0.436) = 0$ and $f'(1.459) = 0$. Both scenarios are unrealistic; however, they allow to compare estimation performances of our method with respect to situations where some prior/oracle information is given. Lastly, we consider a standard Gaussian process regression (GPR) that employs an empirical step to estimate stationary points from the sampled curves, as briefly described below. 

\begin{figure}
	\centering
	\includegraphics[width=.85\textwidth]{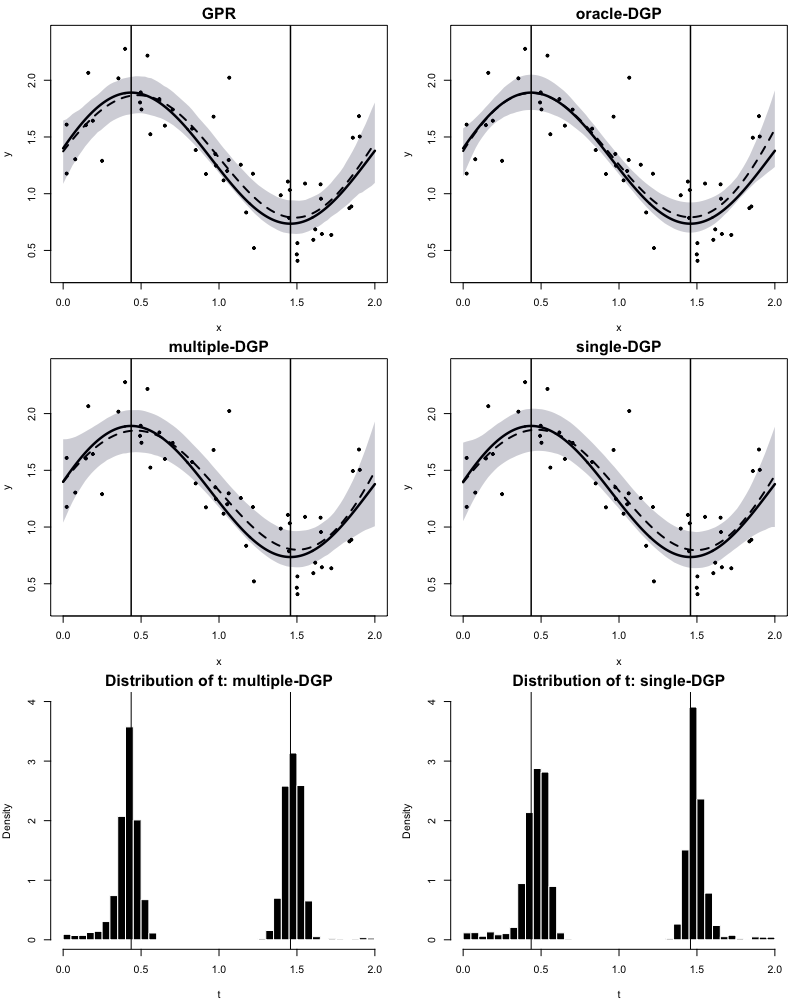}
	\caption{Simulated data. {\it Top two rows:} Estimated curves for one simulated dataset under different models: A standard GPR model and three different DGP models (oracle, multiple and single), as described in the text. The true regression function is shown as a solid line and the estimated curves as dashed lines. Dots indicate input values. {\it Bottom row:} Posterior distributions of $t$, for single and multiple DGPs, with vertical lines indicating the locations of the true stationary points.}
	\label{fig:pred_hist}
\end{figure}

Figure \ref{fig:pred_hist} shows true and estimated regression functions, and corresponding 95\% credible intervals, for one simulated dataset, under oracle, multiple and single DGP and a standard GPR model.  
Although the estimated curves are pretty similar under the different scenarios, the credible intervals of oracle-DGP are narrower, especially where the true function is significantly increasing or decreasing (see for example $x=1$ with $|f'(1)| = 1.74$ and $x=1.45$ with $|f'(1.45)| = 0.047$). This result is of course expected, given that this setting assumes full knowledge of the number and locations of the stationary points.  Multiple-DGP and single-DGP both result in slightly wider credible intervals, due to the uncertainty on the locations of the stationary points. A similar result is obtained with standard GPR.

\begin{figure}
	\centering
	\includegraphics[width=.8\textwidth]{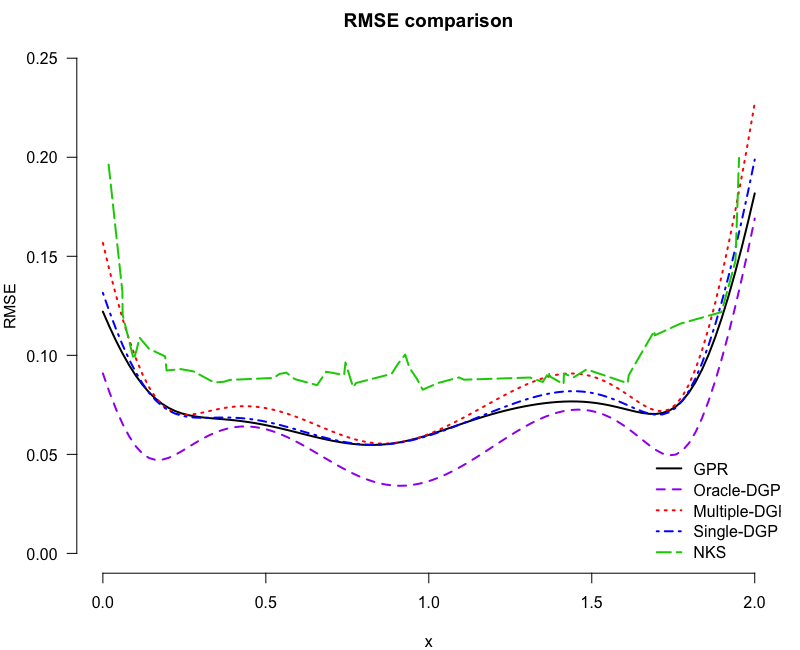}
	\caption{Simulated data. Root mean squared errors (RMSE) between the true regression function and the estimated curves, as a function of $x$, averaged across the 100 simulated datasets, for standard GPR, three different DGP models (oracle, multiple and single), as described in the text, and the NKS method of Song et al. (2006). The GPR and DGP methods use the test input $x_i^*$ while NKS uses the original input $x_i$.}
	\label{fig:rmse_compare}
\end{figure}

In order to further assess estimation performances, we calculated root mean squared errors (RMSE) between the true regression function and the estimated curves, as a function of $x$, averaged across the 100 simulated datasets, as RMSE$(x^*_i)$ = $\sqrt{\frac{\sum_{l=1}^{100}(f(x_i^*) - \hat{f}_{(l)}(x_i^*))^2}{100}}$,
with $\hat{f}_{(l)}(x_i^*)$ the Bayes estimate of $f$ at the test input $x_i^*$ for the $l$-th dataset. Plots of the RMSE under the different settings are shown in Figure \ref{fig:rmse_compare}. In the same figure, as an additional comparison, we report results we obtained by applying a two-step frequentist method proposed by \cite{Song2006} that first employs nonparametric kernel smoothing (NKS) to estimate the curve derivatives and then finds the stationary points as those points at which the first derivative of the curve is zero. Results in Figure \ref{fig:rmse_compare} confirm the superior performance of oracle DGP, with single and multiple DGP performing comparably to standard GPR. Also, DGP methods show generally lower RMSE across $x$ than NKS.

Next, we evaluate performance on the estimation of the stationary points. The bottom row of Figure \ref{fig:pred_hist} shows the posterior distribution of $t$ from multiple-DGP and single-DGP obtained with the MCEM algorithm. Vertical lines show locations of true stationary points. These are correctly identified in both settings, even though some little densities are seen near $x = 0; 2$ resulting in wider credible intervals of $f$ at the boundaries. We note that the posterior distribution of $t$ in single-DGP is reassuringly bimodal and points to two stationary points.  Furthermore,  we remark again that our implementation of multiple-DGP constrains the prior on $t_1,\ldots, t_M$ for identifiability. In particular, we partition the interval using oracle information, with each sub-interval containing one true stationary point.  Single-DGP does not rely on oracle information, and is recommended and generally preferred to multiple-DGP.

Point estimates of the stationary points can be obtained from the posterior distributions of $t$. by calculating HPD intervals and MAP estimates, as explained above. For our single-DGP model, the empirical 95\% HPD intervals averaged over 100 simulated data were $(0.253, 0.555)$ for $\hat{t}_1$ and $(1.325, 1.648)$ for $\hat{t}_2$.  Furthermore, root mean square errors (RMSE) between the true and estimated stationary points, averaged over the 100 simulated datasets as RMSE$(\hat{t}_j)$ = $\sqrt{\frac{\sum_{l=1}^{100}(\hat{t}_j^{(l)} - t_j)^2}{100}}$, with $\hat{t}_j^{(l)}$ the estimate of $t_j$ in the $l$-th simulated data set, were RMSE$(\hat{t}_1)$= {0.0415} and RMSE$(\hat{t}_2)$= {0.0456}, demonstrating that single-DGP with HPD-based estimation provides high estimation accuracy. For a comparison, with NKS we got RMSE$(\hat{t}_1)=0.1410$ and RMSE($\hat{t}_2$)=0.1760. 
Standard GPR does not provide a posterior estimate of $t$, although additional steps can be employed to find stationary points for each posterior samples of $f$, for example by empirically finding the points such that the sign of the difference of differences of $f$ changes. These samples of $t$ have variable dimensions, posing challenges to posterior summary. When implementing such strategy, we calculated HPD regions based on the posterior of $t$ and determined MAP estimates within each interval. With this procedure, we obtained RMSE$(\hat{t}_1)=0.0431$ and RMSE($\hat{t}_2$)=0.0483. We remark that this strategy is sub-optimal as it requires an ad-hoc step to obtain the posterior of $t$.

Additional simulations with different signal-to-noise ratios (SNR) and sample sizes, and sensitivity analyses, are in the Supporting Information.  Results show that, in situations where the true function has some stationary points, providing derivative information does enhance estimation. We also repeat simulations with a Mat\'ern kernel.

\section{ERP Data Analysis}
\label{sec:ERP}
Event-related potentials (ERPs) are electrical potentials that represent electroencephalography (EEG) brain signals recorded in response to specific stimuli. ERP signals measured from EEG experiments have become a common tool in psychological research and neuroscience, due to their ability to provide information about a broad range of cognitive and affective processes. In a typical EEG experiment, to increase signal-to-noise ratio, stimuli are applied repeatedly and the resulting ERP waveforms are averaged across multiple trials for each subject and, often, across multiple subjects \citep{Luck2012}. A primary focus of ERPs analysis is to identify characteristic ``peaks" and ``dips" of the curve and their latency, i.e. position in time (in milliseconds, ms)  and/or magnitude (in microvolts), see top left plot of Figure \ref{fig:ERP_1}.

According to current best practices in ERP methodology, the identification of components is achieved by visual inspection of the grand-averaged  ERP curve (average of all the subjects's data) and constrained by previous findings in the literature using similar experimental designs \citep{Luck2005,Luck2012}. Then, the peak latency and peak amplitude for that component are estimated within each subject.  This is achieved by within subject averaging and then component estimation.  Component latency is estimated first, as the point where the within subject curve reaches maximum amplitude, or where the component reaches half the area under the curve, and peak amplitude is estimated next as the average amplitude at latency.   When latency is not used, amplitude is typically estimated as the average amplitude across an entire component, with all points in the interval of the component contributing equally.  This is, however, unwise, since the value of interest is often the magnitude of the response, but uncertainty on how to measure this makes it necessary to use a broader interval to ensure all of the component is captured.   
This discussion highlights the need for quantitative methods that can identify characteristic components and produce estimates of their latency. In the analysis below we show how our proposed method can be used to estimate smooth ERP curves and automatically identify peaks and their latencies, through the detection of the stationary points of the curves. Unlike empirical techniques, our model-based approach also provides uncertainty quantification on estimated quantities, in the form of HPD intervals. 

\subsection{Experimental study}
\begin{figure}
	\centering
\includegraphics[width=.45\textwidth]{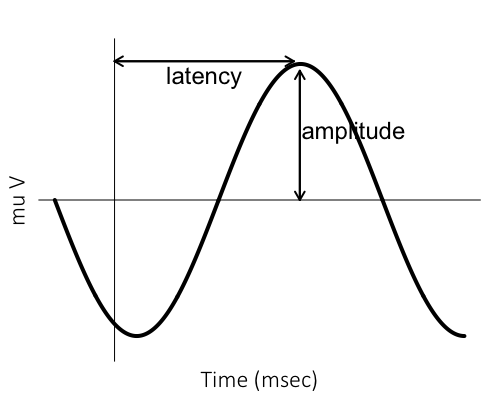}	
\includegraphics[width=.45\textwidth]{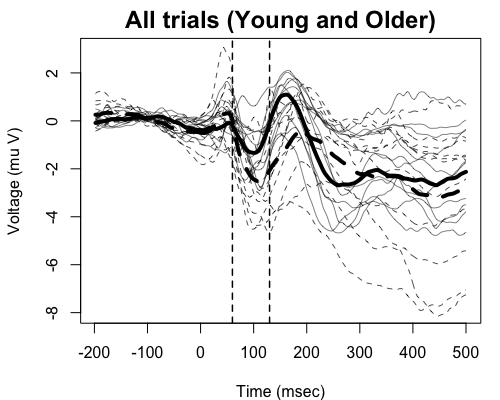}\\
\includegraphics[width=.45\textwidth]{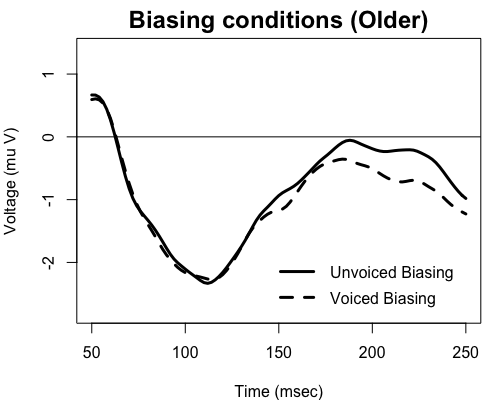}
\includegraphics[width=.45\textwidth]{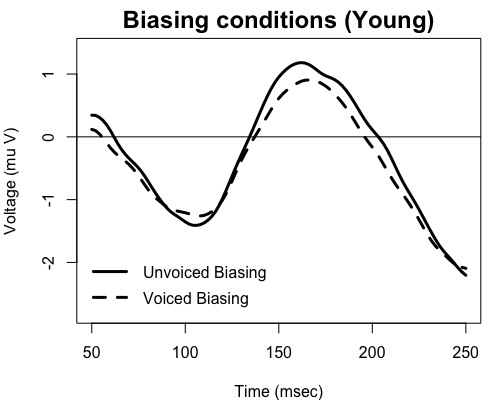}
	\caption{ERP Data Analysis: {\it Top left:} Illustration of amplitude and latency of characteristic components of an ERP signal. {\it Top right}: Subject-level ERP waveforms, averaged over all trial conditions and 6 electrodes. The continuous and dashed thick curves are the benchmark ERP averaged across young and older subjects, respectively. The 0ms time, which corresponds to time point 100, is the start of the onset of sound. Points 101 - 350 represent the time the stimulus is played. The N100 component of interest is the dip characterizing the signal in the time window between the vertical dashed lines. {\it Bottom plots}: ERP waveforms averaged across older and young subjects, for the two (voiced vs unvoiced) lexical biasing trial conditions of interest.} 
	\label{fig:ERP_1}
\end{figure}

We use data from an experiment performed at Rice University on speech recognition. Our ability to recognize speech from a complex acoustic signal depends on merging bottom-up sensory information with top-down expectations about what we are hearing.  The goal of the experiment was to determine whether or not early stages of speech perception are independent of top down influences. In our analysis we consider data from two lexical biasing trial conditions (voiced vs unvoiced), with the aim of investigating the effects of lexicality on phoneme identification, specifically whether the identification is biased towards phonemes which form familiar words. Each experiment yield a total of 2304 trials and EEG was continuously recorded during the task. More details on the experimental design and the pre-processing of the EEG signals are in the Supporting Information. In this paper, we analyze data from 11 college-age students and 11 older controls. Difficulties in perceiving speech, especially in noisy environments, is common with aging \citep{Peelle2016} and might be attributable to a general slowing of the cortical processes. ERP components associated with speech perception are expected to have longer latencies in older adults than in younger adults \citep{Tremblay2003}.  Figure \ref{fig:ERP_1} shows the ERP waveforms for each of the subjects, averaged over all trial conditions and the 6 electrodes. 

Typical analyses of ERP data average ERP waveforms over a window of interest and a given condition to calculate magnitudes and latencies of specific components. For example, \cite{Noe2019} argue for the so called N100 component to be of interest, as this captures phonological (syllable) representation. This component is identified by the latency of the negative deflection, or dip, that typically characterizes the ERP signal in the time interval [60, 130]ms, after the onset of the stimulus (see Figure \ref{fig:ERP_1}). 
For our analysis, we considered a larger time window that comprises the interval [50, 250]ms, containing 101 observation points, to include a second ERP feature of interest, the so called P200, an auditory component that represents some aspect of higher-order perceptual processing, modulated by attention, and that typically peaks at around 200ms. Recent electrophysiological investigations using other imaging modalities have identified the time window around the P200 as potentially critical for processing higher-order speech perception processes \citep{Leonard2016}. Given the goal of the study, and the P200’s potential role in speech perception processes, this is also an ERP component of interest.
The ERP waveforms for the two (voiced vs unvoiced) lexical biasing trial conditions, averaged across young and older subjects, are shown in the lower plots of Figure \ref{fig:ERP_1}. Although averaging across subjects leads to smoother ERP waveforms, the subject-specific variability observed in the subject-level ERP waveforms is washed out, and the information about subject-level ERP is lost. 

\subsection{Results}
Given that we have available subject-level data on multiple subjects, we fit our GP regression model as $y_{i, s} = f_s(x_i) + \epsilon_{i, s},$ with $\epsilon_{i, s} \sim N(0, \sigma^2)$ and $f_s(\cdot) \sim GP(0, k(\cdot, \cdot; \tau, h))$, subject to $f_s'(t_{m_s, s}) = 0$ for $m_s = 1, \dots, M_s$,
with $s=1,\ldots, S$ indicating the subject, and $i=1,\ldots, n$ the time points. We assume that subjects share the same set of time points $\{x_i\}_{i = 1}^n$ and use the SE covariance kernel with common parameters $(\tau, h)$, while allowing subject-level stationary points $t_{m_s, s}$. Within each group of subjects, older and young, we assume a common error variance $\sigma^2$ with an IG prior specified to match the empirical moments calculated based on the data on all subjects. For the prior of $t$, we use our knowledge that the occurrence of the N100 and P200 components is likely in the middle part of the chosen interval and specify a Beta(3, 3) prior. When such knowledge is available, a Beta distribution is preferred over the uniform as it avoids a potentially inflated density at the end points of the time interval. 

\begin{figure}
	\centering
	\includegraphics[width=.4\textwidth]{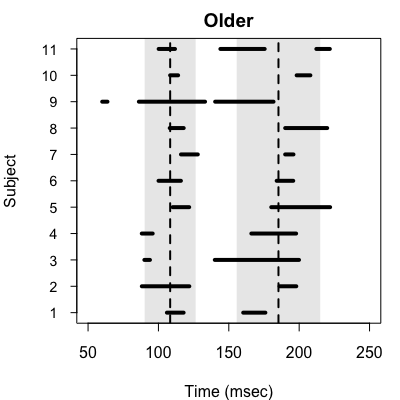}
	\includegraphics[width=.4\textwidth]{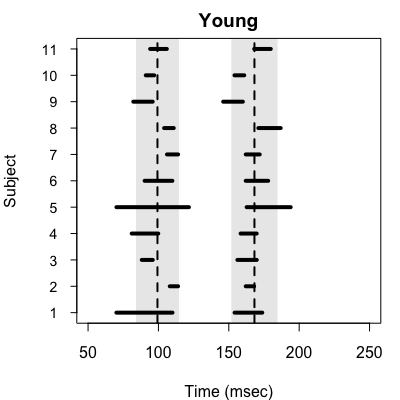}\\
	\includegraphics[width=.4\textwidth]{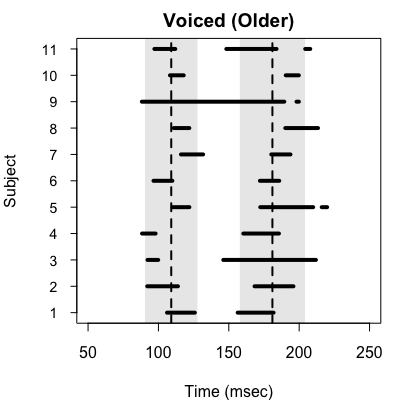}
	\includegraphics[width=.4\textwidth]{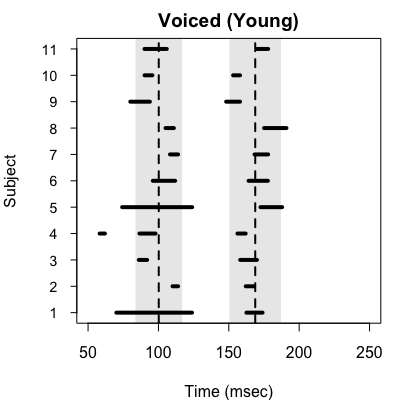}
	\includegraphics[width=.4\textwidth]{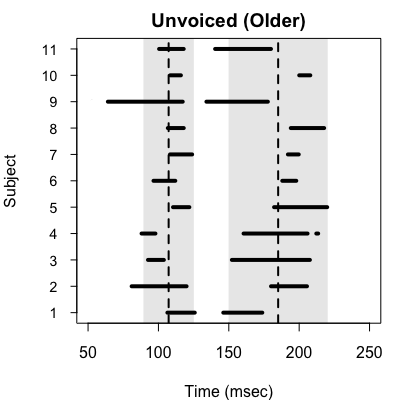}
	\includegraphics[width=.4\textwidth]{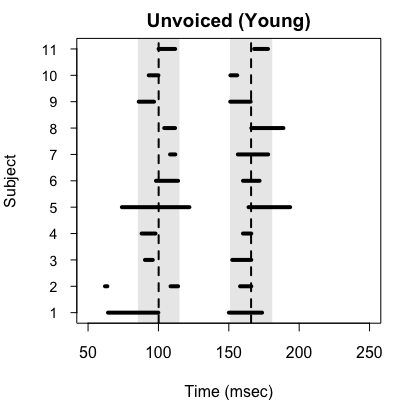}
\caption{ERP Data Analysis:  {\it Left column:} 95\% HPD regions of the posterior distributions of $t$ for the older group, for all subjects and for the voiced and unvoiced biasing trial conditions. {\it Right column:} 95\% HPDs regions of the posterior distributions of $t$ for the young subjects group. In all plots, the dash vertical lines indicate the posterior means obtained by fitting a Gaussian mixture to the posterior samples of $t$,  averaged across subjects, and the 95\% CI calculated as (mean $\pm$ 1.96 std) and shown as shaded areas. }
	\label{fig:ERP_2}
\end{figure}

We show results when fitting our model to the ERP subject-level data averaged over trials and electrodes, as depicted in Figure \ref{fig:ERP_1}. In our analyses, we considered older and young subjects separately. Furthermore, for each group,  separate analyses were performed by considering all the data first (see ERP waveform data depicted in the top plot of Figure \ref{fig:ERP_1}), and then by separating voiced and unvoiced biasing trial conditions  (see bottom plots of Figure \ref{fig:ERP_1}).  Performing separate analyses allows us to gain an understanding of how early cortical responses to speech stimuli change with age and under different conditions.  Figure \ref{fig:ERP_2}  shows the 95\% HPD regions of the posterior distributions of $t$. Plots show substantial subject-level variability, which would be lost when averaging data across subjects, as typically done in ERP data analyses.  The observed variation in the location of the stationary points, i.e. the latencies of ERP components, clearly indicates subject-specific effects. Also, few subjects show HPDs with multiple intervals, suggesting possible outliers.

\begin{table}
\begin{center}
\begin{tabular}{lcc}
\hline
Group& $\hat t_1$ & $\hat t_2$ \\ 
\hline
young                   & 99.31 (7.78)    & 168.22 (8.39)  \\
voiced (young)       & 100.23 (8.46)    & 168.79 (9.34) \\
unvoiced (young) & 100.16 (7.55)    & 165.84 (7.67)    \\
older                  & 108.34 (9.25) & 185.29 (15.15)\\
voiced (older)    & 109.11 (9.51) &  180.99 (11.82)  \\
unvoiced (older)& 107.26 (9.15) & 185.07 (17.98) \\
\hline\\
\end{tabular}
\end{center}
\caption{ERP Data Analysis: Estimated mean components, from a Gaussian mixture model with two components fitted to the posterior samples of $t$,  averaged across subjects, with averaged standard deviations in parentheses.}
\label{tab:one}
\end{table}

As a further validation, we used the R package {\it Rmixmod} to fit two-component Gaussian mixtures to the posterior samples of $t$ obtained from the MCEM algorithm. We set a maximum of 50 iterations and used several runs of the EM algorithm with random initialization, each stopping when the log-likelihood increment is smaller than 0.0001. Table \ref{tab:one} reports the estimated mean components,  averaged across subjects, with averaged standard deviations. Averaged means and 95\% CIs, calculated as (mean $\pm$ 1.96 std), with std indicating standard deviation, are also reported in Figure \ref{fig:ERP_2}. Results show clear latency differences between older and younger adults, with the N100 peak slowed by approximately 10ms and P200 by 22ms. This supports the claim that some differences between older and younger adults in speech perception can be attributed to the speed of cortical processing of speech sounds. Our results also show greater variability in the timing of ERP components among older adults than younger adults for the P200 components. These differences can have enormous consequences on conclusions that we can draw from the standard way of analyzing ERP data. For example, since older adults are much more variable in their speech perception abilities than younger adults, this approach could be used to identify latencies of different components at the subject level and relate those latencies to behavioral differences.  
Overall, our work clearly shows the advantages of model-based approaches for ERP data that can identify subject-specific components and their latencies in a non-subjective, quantitative way.  

\section{Discussion}
\label{sec:disc}
We have proposed a semiparametric Bayesian model to efficiently infer the locations of stationary points of an unknown function. The approach uses Gaussian processes as flexible priors while imposing derivative constraints to control the function's shape via conditioning. We have developed an inferential strategy that intentionally restricts estimation to the case of at least one stationary point, bypassing prior specifications on the number of stationary points and avoiding the varying dimension problem and its computational complexity. 
 
We have illustrated the proposed methods using simulations and an analysis of ERP data. By applying our approach to data on younger and older adults during a speech perception task, we have demonstrated how the time course of speech perception processes changes with age. Our results have clearly shown the advantage of model-based approaches to ERP data analysis over traditional methods that arbitrarily average data across subjects.  In the case study, we reported results on the estimation of latency, as this was the primary objective. However,  peak amplitudes and other quantities can be obtained from the estimated stationary points and latencies, unlike current practices that use empirically chosen windows around a peak identified from globally averaged data. Also, extensions to hierarchical multi-subject models with subject-level random effects can be of practical relevance. 

In our data analysis, we have averaged EEG data across trials and across multiple electrodes, to obtain smooth subject-level ERP signals. Posterior predictive checks (in the Supporting Information) do not suggest inadequacy of our model for such data. 
Although this is reassuring, the iid assumption on the error term of our model
can still be restrictive for ERP signals, particularly the lack of flexibility to account for autocorrelation. We are currently investigating modeling choices that use autocorrelated errors to mitigate this limitation, which also lead to more challenging scenarios when fitting the model. Furthermore, extensions accounting for a design matrix would allow to evaluate ``2-way interactions", including age and conditions, instead of performing separate analysis on groups of data. Finally, while having common GP parameters across subjects helps borrowing information and avoid possible overfitting, extensions that use subject-dependent choices could also be considered,  even though this would make the inferential procedure substantially more complicated.

\bibliographystyle{biom}
\bibliography{bgp_derivative}

\section*{Supporting Information}
MCEM Algorithm, Tables, and Figures referenced in Sections 3,4 and 5 are available with this paper at the Biometrics website on Wiley Online Library. Behavioral and ERP data are published on PsyArxiv (DOI: 10.17605/OSF.IO/C7K4S) at https://osf.io/c7k4s/. R code to recreate results is at https://github.com/chenghanyustats/DGP-MCEM. 

\end{document}